\documentclass[journal=jpclcd,manuscript=letter]{achemso}

\SectionNumbersOn
\usepackage{amsmath}
\usepackage[version=3]{mhchem} % Formula subscripts using \ce{}
\usepackage[font=small,labelfont=bf]{caption}
\usepackage[font=small,labelfont=bf]{subcaption}
\usepackage{threeparttable}
\usepackage{booktabs,siunitx}
\usepackage{physics}
\usepackage{bm}
\usepackage{hyperref}
\usepackage{setspace}
\usepackage{multicol}
\author{Junhan Chen}
\affiliation[Upenn]
{Department of Chemistry, University of Pennsylvania, Philadelphia, Pennsylvania 19104, USA}
\author{Joseph Subotnik}
\email{subotnik@sas.upenn.edu}
\phone{+1 (215) 746-7078}
\affiliation[Upenn]
{Department of Chemistry, University of Pennsylvania, Philadelphia, Pennsylvania 19104, USA}

\title[An \textsf{achemso} demo]
  {Nonadiabatic Potential Energy Surfaces For A Molecule on a Surface as Found by Constrained Complete Active Space Theory}

\abbreviations{IR,NMR,UV}

\begin{document}


\begin{tocentry}
\begin{center}
\includegraphics[width=7.25cm, height=4.45cm]{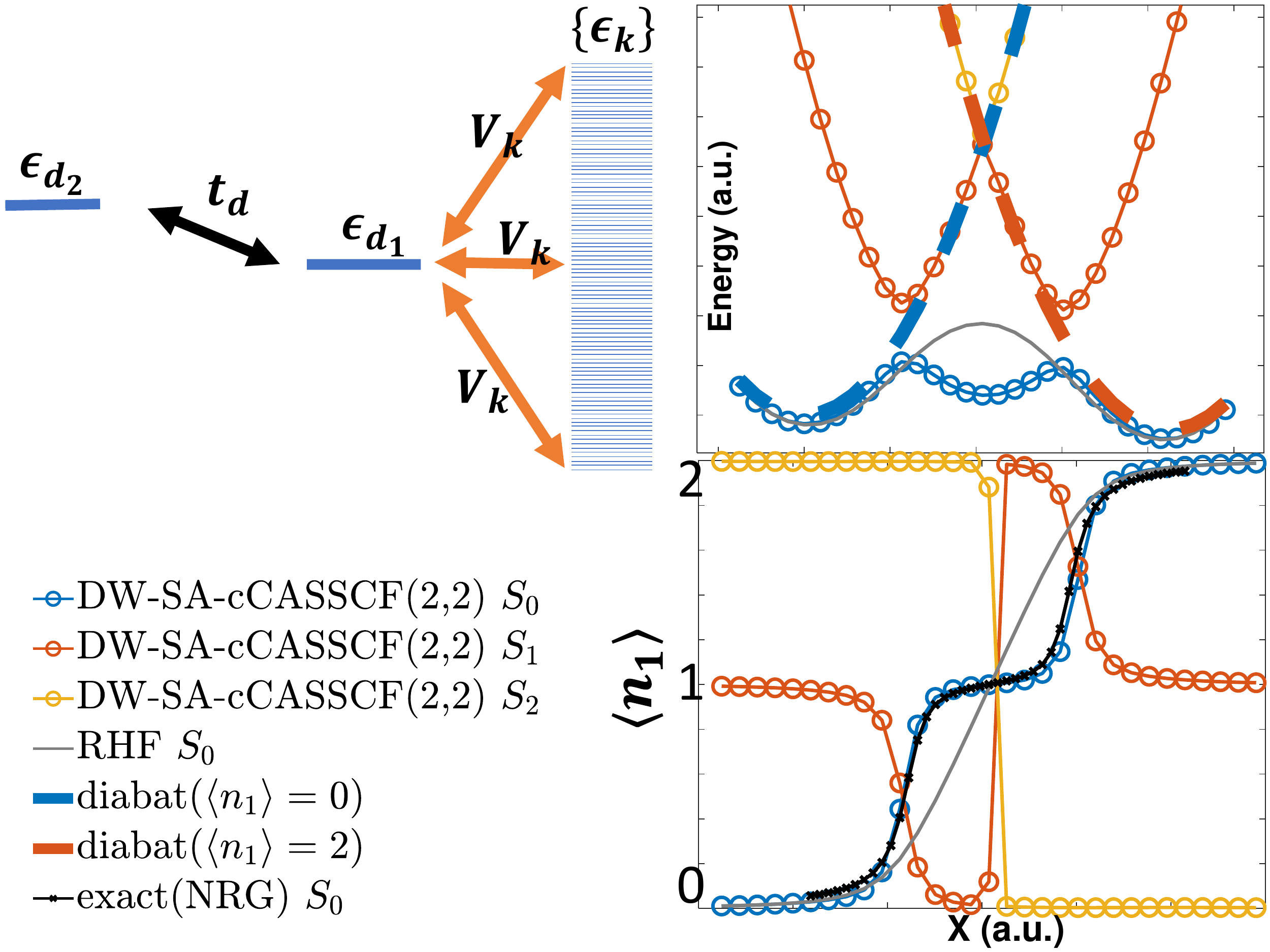}
\end{center}
% Some journals require a graphical entry for the Table of Contents.
% This should be laid out ``print ready'' so that the sizing of the

% Inside the \texttt{tocentry} environment, the font used is Helvetica
% 8\,pt, as required by \emph{Journal of the American Chemical

% The surrounding frame is 9\,cm by 3.5\,cm, which is the maximum
% permitted for  \emph{Journal of the American Chemical Society}
% graphical table of content entries. The box will not resize if the
% content is too big: instead it will overflow the edge of the box.

% This box and the associated title will always be printed on a
% separate page at the end of the document.

\end{tocentry}

\begin{abstract}
  In order to study electron-transfer mediated chemical processes on  a metal surface, one requires not one but two potential energy surfaces (one ground state and one excited state) as in Marcus theory. In this letter, we report that a novel, dynamically-weighted, state-averaged constrained CASSCF(2,2) (DW-SA-cCASSCF(2,2)) can produce such surfaces for the Anderson Impurity model. Both ground and excited state potentials are smooth, they incorporate states with a charge transfer character, and the accuracy of the ground state surface can be verified for some model problems by renormalization group theory. Future development of  gradients and nonadiabatic derivative couplings should allow for the study of non-adiabatic dynamics for molecules near metal surfaces. 
\end{abstract}


\section{Introduction}

A great deal of chemistry occurs at interfaces. One can build 
and run a current through a molecular junction, \cite{alemani2006electric,danilov2006electron,donarini2006dynamical,henningsen2007inducing,datta1997current,samanta1996electronic,nitzan2003electron,mujica1996current,hsu1997sequential,stipe1998single},
chemisorb and dissociate chemical bonds,  \cite{jiang2016electron,maurer2017mode,yin2018dissociative,chen2018vibrational} or scatter molecules  \cite{waldeck1985nonradiative,wodtke2004electronically,bartels2011energy,kandratsenka2018unified,shenvi2009dynamical} as a few examples. At a microscopic level, the physics underlying these phenomena is much richer and complex than what can be found in solution because of the continuum of electronic states. For instance, to model the phenomena above, one must be able to describe molecular resonance effects on metal surfaces \cite{gavnholt2008delta}, electron-hole-pair quenching \cite{jiang2016electron,persson1982electron}, electron-coupled adsorption \cite{bunermann2015electron} and electron-coupled vibrational motion \cite{morin1992vibrational, huang2000vibrational}.

One crucial problem when describing dynamics at metal surfaces is how to model coupled nuclear-electronic motion.  Because of the continuum of electronic states at a metal interface, the Born-Oppenheimer approximation can routinely break down\cite{wodtke2004electronically}
and at present, there are far fewer means to model such dynamics (at least relative to dynamics in solution).  As originally devised by Suhl \cite{d1975brownian}, the simplest approach for going beyond Born-Oppenheimer at a metal interface is to map all non-Born-Opppenheimer effects into a Markovian ``electronic friction'' tensor that damps the {\em nuclear} motion.  This approach has been developed by several researchers over the years \cite{
bohnen1975friction,brandbyge1995electronically,jin2019practical,dou2018perspective}, and non-Markovian electronic friction tensors have also been proposed. The strength of the electronic friction approach is that, in principle, one runs dynamics along the ground state (or a thermally averaged state) and all information about electronic excited states and nonadiabatic transitions is wrapped into the electronic friction tensor (which can be calculated either within a DFT framework [assuming independent particles] \cite{maurer2016ab} or approximated in different ways\cite{rittmeyer2015electronic,juaristi:2008:prl_ldfa}). 

Despite these successes, friction tensors have limitations and have not been able to describe, e.g., the Wodtke charge-transfer  experiments.\cite{bartels2011energy}  To go beyond these approaches, one would like to run more fully nonadiabatic dynamics at a metal surface, either through an IESH framework\cite{shenvi2009nonadiabatic} or a  BCME ansatz\cite{dou2016broadened}; note that IESH has been able to model some of the Wodtke experiments (but not all).  For these more robust dynamical schemes, however,  one must go beyond the ground state and also calculate excited electronic states  \cite{lichten1963resonant,behler2007nonadiabatic}
as well as the coupling between electronic states. 
At this point, the electronic structure becomes a real headache as, near a metal surface, there is a continuum of states and one cannot afford to generate too many states; furthermore, with a small or no band gap, it is not immediately obvious how to choose the relevant states. 

For extended systems, unfortunately, the traditional high accuracy quantum mechanics methods can be impractical, e.g.
 coupled-cluster singles, doubles and full triples (CCSDT) \cite{raghavachari1989fifth,crawford2007introduction}, second-order perturbation with complete active space (CASPT2) \cite{andersson1990p,andersson1992second} and  multireference configuration interaction (MRCI) \cite{bruna1987excited} or full configuration interaction (FCI) \cite{sherrill1999configuration,sherrill1996computational}. If one wishes to  calculate excited states for molecules on small clusters or metal surfaces, the standard approaches might include:

\begin{enumerate}

\item Constrained DFT (CDFT). \cite{wu2006extracting,wu2005direct,behler2007nonadiabatic,kaduk2012constrained,meng2022pragmatic} CDFT has been applied to a few realistic calculations on surface, e.g. charge transport\cite{goldey2017charge,ma2020pycdft} and the energy-level alignment near metal surfaces\cite{souza2013constrained}. However, the method can fail in the case of strong molecule-metal coupling (i.e. strong hybridization) \cite{gavnholt2008delta} and fractional charge transfer \cite{mavros2015communication}.

\item For the specific case of charge transfer, $\Delta$ SCF is a powerful tool. One  runs two calculations, each  with different numbers of electrons.  Previous work has successfully modeled at least one case of a molecular resonance near metal surfaces\cite{gavnholt2008delta} with a modified $\Delta$ SCF called linear-expansion $\Delta$ SCF (and it is believed to be  especially relevant in Newns-Anderson-type\cite{newns1969self,anderson1961localized} systems), but it is not clear how to pick linear-expansion coefficients of Kohn-Sham (KS) orbitals for the charge transfer between a molecule and a metal surface; And the nuclear gradient for excited states is very difficult to get.

\item The GW approximation \cite{hedin1965new,onida2002electronic, liu2019accelerating}.  One can use many-body theory to extract accurate excitation energies for molecules on clusters.  However, these calculations are already expensive and it is difficult to imagine using many-body theory to calculate nonperturbative excited states in such a manner that allows for the calculation of meaningful excited state gradients and derivative couplings.

\item Most powerful of all are embedding methods \cite{kluner2002periodic,lahav2007self,zhu2019efficient,zhu2021ab,chan2011density,knizia2012density}, which combine high-accuracy quantum chemistry methods with DFT. Embedding approaches have been applied to estimate PESs of excited molecules on clusters \cite{mehdaoui2007understanding} and surfaces \cite{martirez2017prediction}, but such calculations remain expensive; excited state gradients and derivative couplings will be very costly and charge transfer excited state character is difficult to model. 

\end{enumerate}

With this background in mind, over the last few years, our goal has been to construct smooth and qualitatively correct potential energy surfaces for a molecule near a metal surface that should enable us  to run non-adiabatic dynamics\cite{jin2021nonadiabatic,dou2018perspective,jin2019practical}. Because the problem is so difficult--we seek just a {\em few}, smooth electronic states that can capture the electronic structure of an interacting {\em continuum} of electrons--one clearly cannot expect quantitative accuracy in all regimes. Ideally, though, one would like a method that can recover both $(i)$ the strongly adiabatic limit (where motion along the ground state is a good approximation) and $(ii)$ the strongly nonadiabatic limit (where diabatic curves should enter, charge transfer is rare, and Marcus theory applies.)  One consequence of this requirement is that the method should work well $(a)$ for molecules that are strongly or weakly hybridized (chemisorbed or physisorbed) to a surface, and $(b)$ for molecules with strong or weak electron-electron repulsion.

To that end, in the present letter, we will show that a good target approach is to use a 
 dynamically-weighted state-averaged constrained-CASSCF(2,2) (DW-SA-cCASSCF(2,2)).
 For the famous Anderson-Holstein problem (with one or two sites), we will show that such an ansatz can recover multiple smooth potential energy surfaces, where the ground state roughly matches exact numerical renormalization group theory results  (for $U>0$) or just direct diagonalization (for $U=0$).  We will also show that, without constraints or dynamical weighting, standard CASSCF(2,2) cannot perform nearly as well; in particular, such methods cannot recover smooth curve crossings for the potential energy surfaces. Although DW-SA-cCASSCF(2,2) does fail when more than two electrons becomes strongly entangled (as found, e.g., for certain regimes of a two-site Anderson Impurity model), the data below demonstrates that the method would appear to be an outstanding framework for propagating nonadiabatic molecular dynamics on metal surfaces in the near future to model charge transfer. 
% \textcolor{red}{Fourth? Take a glance of the implementation zuxin's FSSH-ER with CASSCF(2,2) states?}.

\section{Theory}

When working with extended electronic structure problems, one of the difficulties is extracting accurate (exact) numerical benchmarks. For this reason, when studying embedding problems, one of the most natural Hamiltonians is the Anderson-Holstein (AH) model. Within a second quantized representation, the Hamiltonian can be written as  $\hat{H}_{tot} = \hat{H}_{el} -  \frac{\hbar^2}{2M}\nabla_x^2$, where the electronic Hamiltonian is:
 \begin{equation}
 \begin{aligned}
 \label{eqn:model} 
     &\hat{H}_{el}=\frac{1}{2}m\omega^2x^2+\hat{H}_{one}(x)+\hat{\Pi}\\
     &\hat{H}_{one}(x)=\epsilon_{d_1}(x)\sum_\sigma d_{1\sigma}^{\dagger}d_{1\sigma}+\epsilon_{d_2}(x)\sum_\sigma d_{2\sigma}^{\dagger}d_{2\sigma}\\
     &+t_d\sum_\sigma(d_{1\sigma}^{\dagger}d_{2\sigma}+d_{2\sigma}^{\dagger}d_{1\sigma})\\
     &+ \sum_{k\sigma}\epsilon_{k\sigma}c_{k\sigma}^{\dagger}c_{k\sigma}+\sum_{k\sigma}V_k(d_{1\sigma}^{\dagger}c_{k\sigma}+c_{k\sigma}^{\dagger}d_{1\sigma})\\
     &\hat{\Pi}=U(d_{1\uparrow}^{\dagger}d_{1\uparrow}d_{1\downarrow}^{\dagger}d_{1\downarrow}+d_{2\uparrow}^{\dagger}d_{2\uparrow}d_{2\downarrow}^{\dagger}d_{2\downarrow})\\
     &\epsilon_{d_1}(x)=e_{d_1}-\sqrt{2}gx\\
     &\epsilon_{d_2}(x)=e_{d_2}-\sqrt{2}gx
     \end{aligned}
 \end{equation}
In this paper, we will not consider nuclear motion and will focus on treating the electronic Hamiltonian, $\hat{H}_{el}$;  the parameter $x$ will be a constant that we can vary for each calculation below (the term $\frac{1}{2}m\omega^2x^2$ representing the potential energy of a nuclear vibration associated with the molecule or impurity).  The operators  $\{\hat{d_1}^\dagger,\hat{d_2}^\dagger\}$ create impurity atomic orbitals, the operators $\hat{c_k}^\dagger$ create  bath (metal surface) atomic orbitals, and $\sigma$ 
denotes an electron spin.
$\epsilon_{d_1}(x)$ and  $\epsilon_{d_2}(x)$ are one-electron ionization energies for the impurities (which linearly depend on the impurity nuclear coordinate $x$); $\epsilon_k$ denotes the energy of bath orbital $k$. $t_d$ is the hopping parameter between site 1 and site 2, $U$ represents the on-site coulomb repulsion for the impurity. $V_k$ represents the hybridization between impurity site 1 and the metal bath, and as in the wide band approximation, is characterized by:
\begin{equation}
    % \Gamma = 2\pi V_k^2 \rho\\
    \Gamma=\Gamma(\epsilon)=2\pi\sum_k |V_k|^2\delta(\epsilon-\epsilon_k) ,
\end{equation}
where $\Gamma$ is assumed to be constant through the whole energy spectrum $\epsilon$.

The difficulty in solving the interfacial electronic structure theory with a metallic solid arises from the low-lying electron-hole-pair excitation within the metal, whose excitation energy is negligibly small; for interfacial problems, there is a true continuum of states and there is no gap between the ground and first excited state. This state of affairs contrasts with the case of an isolated molecule, where the potential energy surfaces are usually well-separated for most of nuclear configuration space with only a few and crossings at isolated geometries.  Moreover, as discussed above, for problems with charge transfer, we require both ground and excited states and ideally the capacity to diabatize the two sets of manifolds; alas standard diabatization schemes  \cite{subotnik2008constructing,subotnik2009initial} are not directly applicable in this case.
% Therefore, it's too expensive to run dynamics using traditional configuration interaction methods (e.g. CISD). Instead, we have tried a selective configuration interaction method \cite{chen2021electronic} but the way to pick proper/crucial configurations is not clear for the semi-{\em ab initio} system, a real SF$_6$ molecule with a tight-binding metal. Then we also tried an alternative method, non-orthogonal configuration interaction based on constrained Hartree-Fock wavefunctions (NOCI/CHF). But our results on the Anderson model show its failure in the strong molecule-metal coupling regime \cite{chen2022active}, which supports the statement in the literature\cite{gavnholt2008delta}. To that end, we it would be much easier to run dynamics if one could extract only a few potential energy surfaces that contain the information of chemical reactions (e.g. charge transfer). 

% Modeling the dynamics is hard.
% Big Problem: many states. Diabatizatoin is tough. So far unanswered 

%Veecently, we worked with a method, called partially-optimized closed-or-open-shell Hartree-Fock (poCOOS-HF)\cite{chen2022active}. The ground state looks good as compared to the exact NRG, but it couldn't get excited states ( and there are still some small discontinuites). Then we further relaxed the core orbitals and worked with a fully optimized closed-or-open-shell Hartree-Fock (foCOOS-HF), which turns out to be a subset of CASSCF(2,2).

With this background in mind, we have sought to explore whether standard (but slightly tweeked)
quantum chemistry methods can solve such interface problem, in particular CASSCF-like solutions. Our recent forays into different versions of Hartree-Fock theory with non-orthogonal orbitals [partially-optimized closed-or-open-shell Hartree-Fock (poCOOS-HF)\cite{chen2022active}]
have pointed to CASSCF(2,2) as the most natural (and improved) starting point.
To that end, we will explore four different CASSCF-based approaches for modeling the AH model:

\begin{enumerate}
\item Standard complete active space self-consistent-field (CASSCF). We will show that this approach works well for capturing static correlation and an accurate ground state energy. However, not surprisingly for the CASSCF experts,  the method can yield incorrect ground-excited state gaps/crossings  with discontinuous surfaces.

\item Dynamically weighted state-averaged CASSCF (DW-SA-CASSCF). State-averaging can fix many of the discontinuities of CASSCF, though at the expense of less accurate ground state energies (again, not surprising to a CASSCF expert). In practice, dynamical-weighting (DW)  usually outperforms static weighting as far as achieving a  better balance between accurate energies and continuous surfaces; dynamic weighting can
avoid some of the usual pitfalls of static state-averaged CASSCF (e.g., example,   the unnecessary discontinuities in the excited state energy surfaces as found for butadiene twisting\cite{glover2014communication}; or the incorrect avoided crossing  geometries as found with Li-F \cite{battaglia2020extended}).
  Our dynamical weighting scheme follows  Ref.  \cite{battaglia2020extended}.    Unfortunately, despite these improvements, we will find that  DW-SA-CASSCF can still fail for the AH model because one cannot always isolate a consistent set of electronic states with meaningful impurity character; the intruder problem remains a nightmare.

\item Constrained CASSCF (cCASSCF). Our experience above will lead us to the concept of a {\em constrained} CASSCF (cCASSCF) optimization procedure, whereby we insist that the molecular population in the active space (spanned by active orbitals $\left\{ t,u\right\}$) always has magnitude equal to unity:
\begin{align}
\label{eqn:constraint1}
    \sum_{\mu\in\textbf{impurity}}\mel{t}{d_\mu^{\dagger}d_\mu}{t}+\mel{u}{d_\mu^{\dagger}d_\mu}{u}=1
\end{align}
This approach helps with the intruder state problem, but excited states can still be discontinuous far away from the crossing regime.

\item Dynamically Weighted State-averaged constrained CASSCF (DW-SA-cCASSCF).  Our best overall candidate for exploring excited state dynamics at an interface, with accurate electronic energies and smooth surfaces, combines both the dynamical weighting and the impurity constraint to definitively solve the intruder state problem for the AH model.

% The following two different constraints are briefly discussed in the letter and we believe the first one is better.
% \begin{eqnarray}
% \sum_{\mu\in\textbf{impurity}}\mel{t}{d_\mu^{\dagger}d_\mu}{t}+\mel{u}{d_\mu^{\dagger}d_\mu}{u}=1\\
% \sum_{\mu\in\textbf{impurity}}\mel{t}{d_\mu^{\dagger}d_\mu}{t}=1 ;  \sum_{\mu\in\textbf{impurity}}\mel{u}{d_\mu^{\dagger}d_\mu}{u}=0
% \end{eqnarray}

\end{enumerate}

For a complete description of the relevant theory and all of the relevant equations needed to define and implement these methods, please see Ref. \cite{chenunpublish}.

\section{Results}

As discussed above, a meaningful electronic structure description of a molecular dynamics on  a metal surface surface should be accurate both in the weak and strong coupling limits, and in the limits of strong and weak electron-electron repulsion.  Because we do not run dynamics here, we cannot check the first condition ($\Gamma/\omega$ big or small). However, we can check the second condition ($U/\Gamma$). Below, we will work with a manifold of electronic states for which the wide-band approximation holds. An  investigation of  different values for U indicates that DW-SA-cCASSCF(2,2) has a great deal of promise.

\subsection{$U =0$, one-site Hamiltonian}
\begin{figure}[htbp]
\centering
\hspace*{-0mm}\includegraphics[width=1\linewidth]{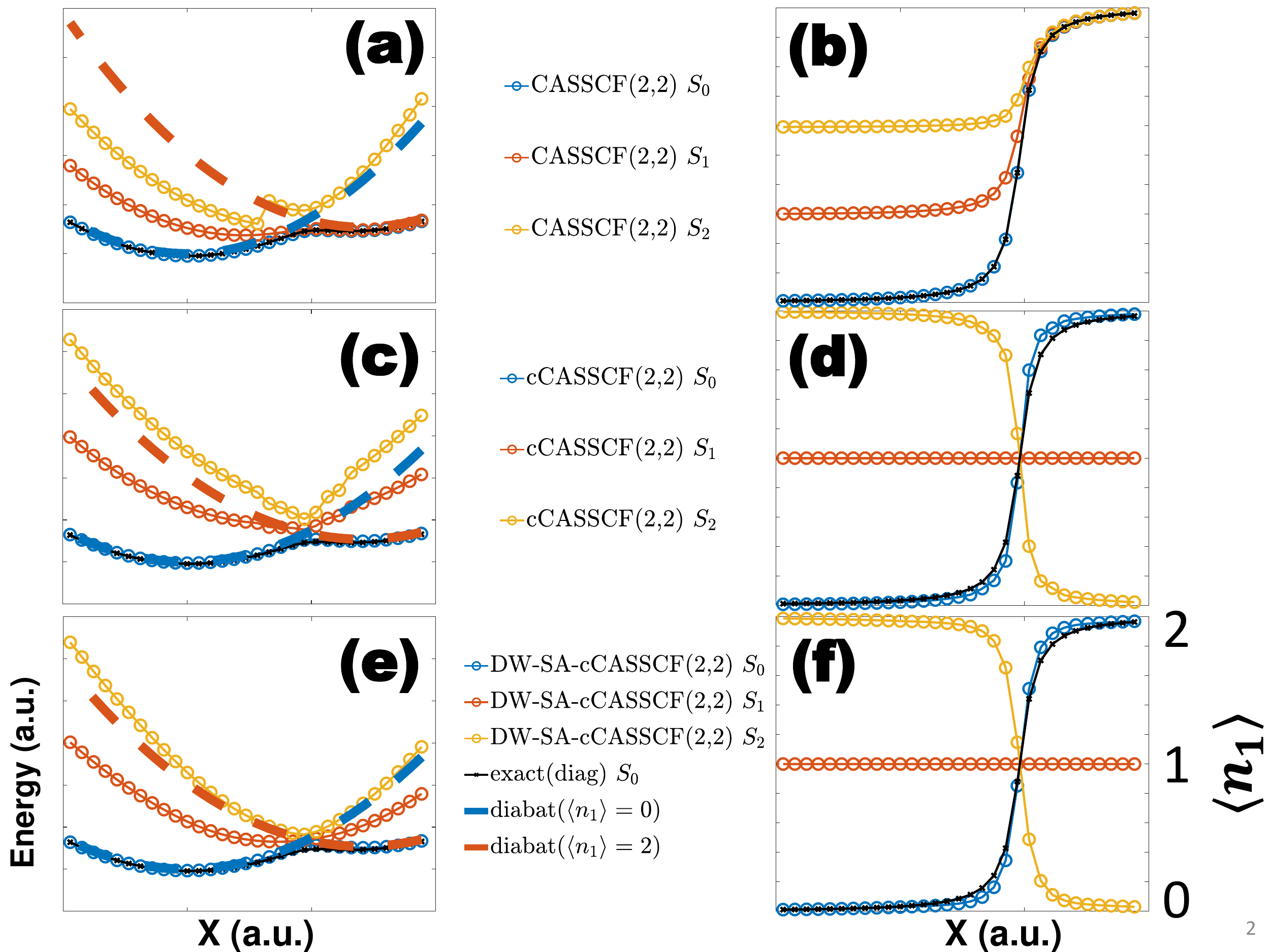}
\caption{The energies and the impurity electron population ($\expval{n_1}$) of the 1-site Anderson impurity model {\em without} electron-electron repulsion for three CASSCF(2,2) methods: (a,b) CASSCF(2,2); (c,d) Constrained CASSCF(2,2) (cCASSCF(2,2)); (e,f) Dynamical-weighted state-averaged constrained CASSCF(2,2) (DW-SA-cCASSCF(2,2)) with $\zeta=40$. On the right hand side, the black line is the exact adiabatic ground state energy and electron populations (by direct diagonalization) and two dashed lines represent two exact diabatic energies with $\expval{n_1}=0$ and $\expval{n_1}=2$. Note that CASSCF(2,2) and cCASSCF(2,2) excited state energies do not agree in  (a,c) because there are so many states around to choose from. The parameter set is $m\omega^2=0.003, g=0.0075, e_{d_1}=0.05, \Gamma=0.01, U=0$ with 101 metal states evenly distributed within the energy window $[-0.05, 0.05]$.}
\label{fig: 1site_casscf1}
\end{figure}

We begin with the case $U =0$, for which an exact solution can be constructed (as there is no electron-electron correlation). 
%Six subfigures in Fig. \ref{fig: 1site_casscf1} are  organised as follows: First, in Figs. \ref{fig: 1site_casscf1}(a-b), we discuss the results of the state-specific CASSCF(2,2) method. Second, in Figs. \ref{fig: 1site_casscf1}(c-d), we address the improvement of the excited states results by the constrained CASSCF(2,2) method and point out its defect of multiple solutions for the excited states. Last, in Figs. \ref{fig: 1site_casscf1}(e-f), we introduce the dynamically-weighted constrained CASSCF(2,2) method, the most out-performed method among the three CASSCF(2,2) methods. 
In Figs. \ref{fig: 1site_casscf1}(a,b), we plot the energy and the electron population as a function of the nuclear coordinate $x$ for traditional state-specific CASSCF(2,2). We find immediately that the method generates three nearly degenerate states. One would hypothesize that\ the two active orbitals do not have any impurity character and the excited states have metal-to-metal excitation character on top of the ground state.  This suspicion is confirmed in Fig. \ref{fig: 1site_casscf1}(b) showing that the electron population behavior for the ground and excited states are nearly the same. 

Clearly,  state-specific CASSCF(2,2) cannot consistently generate the two active orbitals that have any impurity character and thus cannot model charge transfer. To avoid the scenario of Figs. \ref{fig: 1site_casscf1}(a,b), the 2 active orbitals $t,u$ should be a mixture of impurity atomic orbitals $\{d_\mu\}$ and bath atomic orbitals $\{b_\nu\}$ (and not just bath orbitals). This conclusion leads to the constraint in Eq. \ref{eqn:constraint1}. In Ref. \cite{chenunpublish}, we derive in detail the necessary equations that must be implemented in order to solve CASSCF(2,2) with such a constraint.  In short, one can integrate a standard CASSCF routine within a Lagrange multiplier self-consistently loop.

In Figs. \ref{fig: 1site_casscf1}(c,d), we plot results for constrained CASSCF(2,2). As one would hope, now we do see the desired electron population behavior (three curves crossing) in Fig. \ref{fig: 1site_casscf1}(d). However, note that the constrained CASSCF(2,2)  excited state energies do have a small discontinuity; moreover, these energies not ever reach the proper diabatic limit when the impurity is fully occupied $(\expval{n_1}=2)$ or totally unoccupied $(\expval{n_1}=0)$. At bottom, the problem is that, if one optimizes only the ground state energy (and pays no attention to the excited state energies), one risks having inaccurate excited states (a standard feature of CASSCF solutions). Moreover, for this specific problem ($U=0$), no active space is needed for exact diagonalization, and so the excited state energies are not even well defined. Visually, this failure becomes most accute  far away from the curve crossing, where the $S_2$ excited state energy  is quite different from the diabatic energy $diabat(\expval{n_1}=2)$ on the right hand side and the diabatic energy $diabat(\expval{n_1}=0)$ on the left hand side. 

% Now let's be very clear about why the constrained ground-state-specific CASSCF(2,2) cannot give a meaningful excited state under the simplest model: one-site Anderson model {\em without} electron-electron repulsion. Consider the limiting case where the ground state has the impurity site fully occupied, i.e. $\expval{d^{\dagger}d}=\expval{\bar{d}^{\dagger}\bar{d}}=1$. Recall that the general CASSCF(2,2) wavefunction is written as (ignoring the core electrons configuration): 
% \begin{equation}
% \ket{\Psi}^{CASSCF(2,2)}=\alpha\ket{t\bar{t}}+\beta\ket{u\bar{u}}+\gamma\ket{t\bar{u}+u\bar{t}}
% \end{equation}
% Clearly, $(\alpha,\beta,\gamma)=(1,0,0)$ and $t=d,u\in\text{bath}$ (i.e. $\ket{\Psi}^{CASSCF(2,2)}=\ket{d\bar{d}}$) is the solution of the ground state CASSCF(2,2) satisfying the constraint in eq. \ref{eqn:constraint}. But $u$ is not unique: $u$ can be any bath orbitals. In such case, one cannot extract a meaningful excited state from the constrained CASSCF(2,2) (even though these three states represent different charge states on the impurity). Hence, to obtain a meaningful excited state, one must do state-average to minimize the energy summation of the ground state and excited states. 

Finally, in Figs. \ref{fig: 1site_casscf1}(e,f), we plot results for  dynamically-weighted state-averaged constrained CASSCF(2,2). This method appears to give the best of both worlds: an accurate ground states and smooth excited states with the right asymptotic limits.  The method succeeds by smoothly interpolating the fock operator between state-specific and state-averaged regime, while still enforcing the constraint that the impurity must be involved.

%  An exponential form of the weighting factor can be written as: 
%  \begin{equation}
%      \begin{aligned}
%         w_0&=\frac{1}{1+e^{-\zeta (E_1-E_0)}+e^{-\zeta (E_2-E0)}}\\
%         w_1&=\frac{e^{-\zeta (E_1-E_0)}}{1+e^{-\zeta (E_1-E_0)}+e^{-\zeta (E_2-E0)}}\\
%         w_2&=\frac{e^{-\zeta (E_2-E0)}}{1+e^{-\zeta (E_1-E_0)}+e^{-\zeta (E_2-E0)}}
%     \end{aligned}
%  \end{equation},
% where $E_0,E_1,E_2$ are energies of three CASSCF(2,2) states and $\zeta$ is a parameter to control the mixing strength of the ground state and the excited states. When $\zeta \to 0$, it goes back to three-states-averaged CASSCF(2,2) with equal weighting $w_I=\frac{1}{3}$; When $\zeta \to \infty$, it becomes state-specific CASSCF(2,2). In Fig. \ref{fig: 1site_casscf1}, we find that $\zeta=40$ is the best choice.

\subsection{$U >0$ , one-site and two-site Hamiltonian}
Having explored three CASSCF(2,2) methods for 1-site Anderson model {\em without} the electron-electron repulsion, we will now address the 1-site and 2-site Anderson models {\em with}  electron-electron repulsion, which is more realistic and more difficult to solve. 

\subsubsection{One Site}
\begin{figure}[H]
\centering
\hspace*{-0mm}\includegraphics[width=1\linewidth]{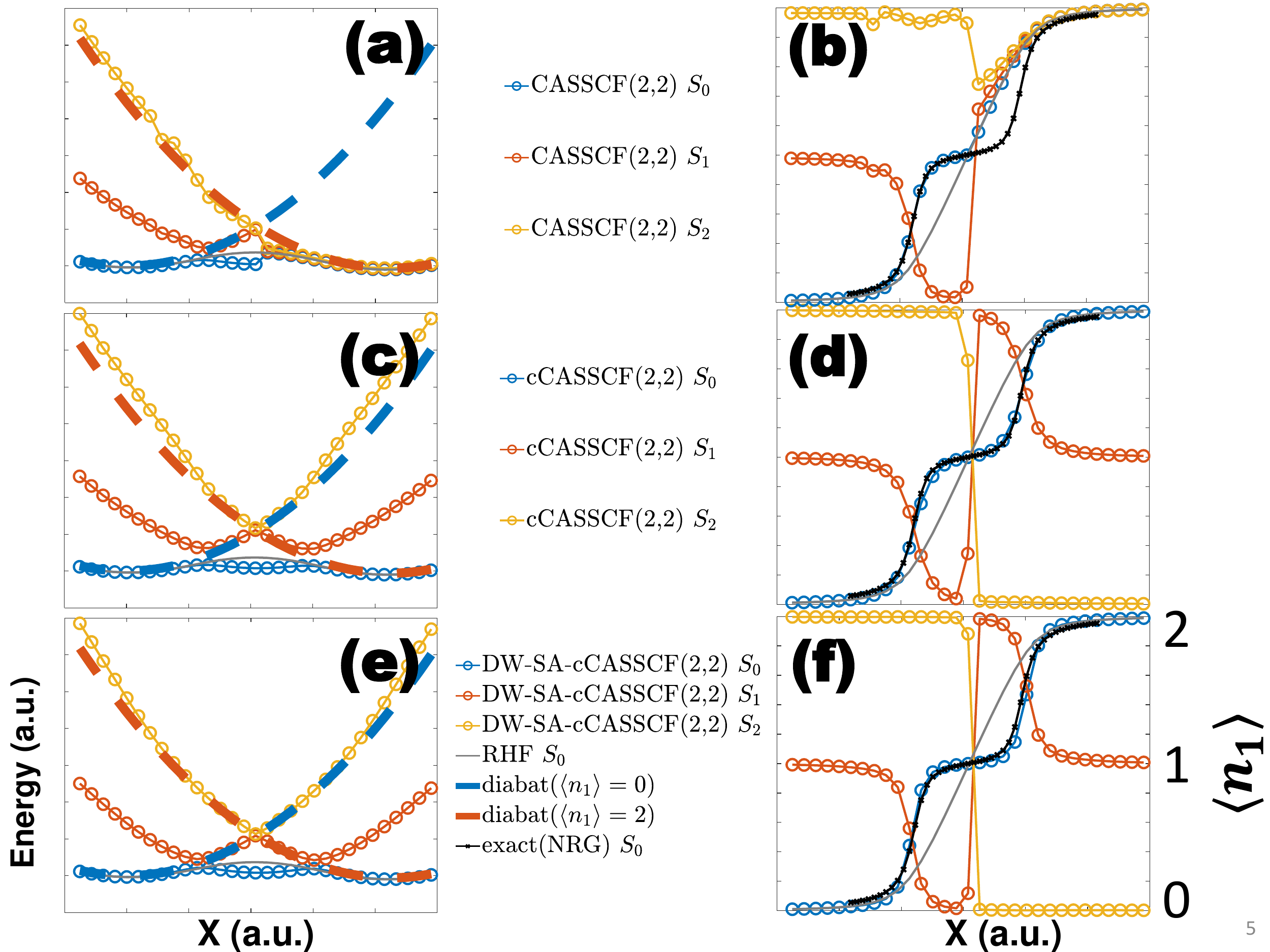}
\caption{The energies and the impurity electron population ($\expval{n_1}$) of the 1-site Anderson impurity model {\em with} electron-electron repulsion for three CASSCF(2,2) methods: (a,b) CASSCF(2,2); (c,d) Constrained CASSCF(2,2) (cCASSCF(2,2)); (e,f) Dynamical-weighted state-averaged constrained CASSCF(2,2) (DW-SA-cCASSCF(2,2)) with $\zeta=40$. On the right hand side, the black line is the exact adiabatic ground state electron population (calculated by numerical renormalization group, NRG) and two dashed lines represent two exact diabatic energies with $\expval{n_1}=0$ and $\expval{n_1}=2$. The blue line without the circle is the RHF energy and electron populations. Note that {\em with} the electron-electron repulsion, DW-SA-cCASSCF(2,2) can produce a triple energy well corresponding to $\expval{n_1}=0, \expval{n_1}=1, \expval{n_1}=2$ while RHF is qualitatively wrong.
The parameter set is $m\omega^2=0.001, g=0.0075, e_{d_1}=0.06, \Gamma=0.01, U=0.1$ with 101 metal states evenly distributed within the energy window $[-0.05, 0.05]$.}
\label{fig: 1site_casscf2}
\end{figure}

In the Fig. \ref{fig: 1site_casscf2}, we plot  results  for the 1-site model {\em with} the electron-electron repulsion
according to the CASSCF(2,2) methods discussed above. Although one cannot recover the exact ground state energy for this entangled Hamlitonian, one can extract the exact impurity populuation using numerical renormalization group theory (NRG).\cite{bulla2008numerical, dou2017born}

According  to Fig. \ref{fig: 1site_casscf2}, our conclusions with $U>0$ are largely the same as our conclusions with $U =0$. In both cases,  DW-SA-cCASSCF(2,2) would appear to offer the best compromise between an accurate ground state and smooth excited states with the correct asymptotic limits. Note that, when $U>0$ (unlike when $U=0$), we find not one but two excited state crossings as the zero-to-one and one-to-two electron transitions are no longer on top of each other.
\subsubsection{Two Sites}
\begin{figure}[H]
\centering
\hspace*{-0mm}\includegraphics[width=1\linewidth]{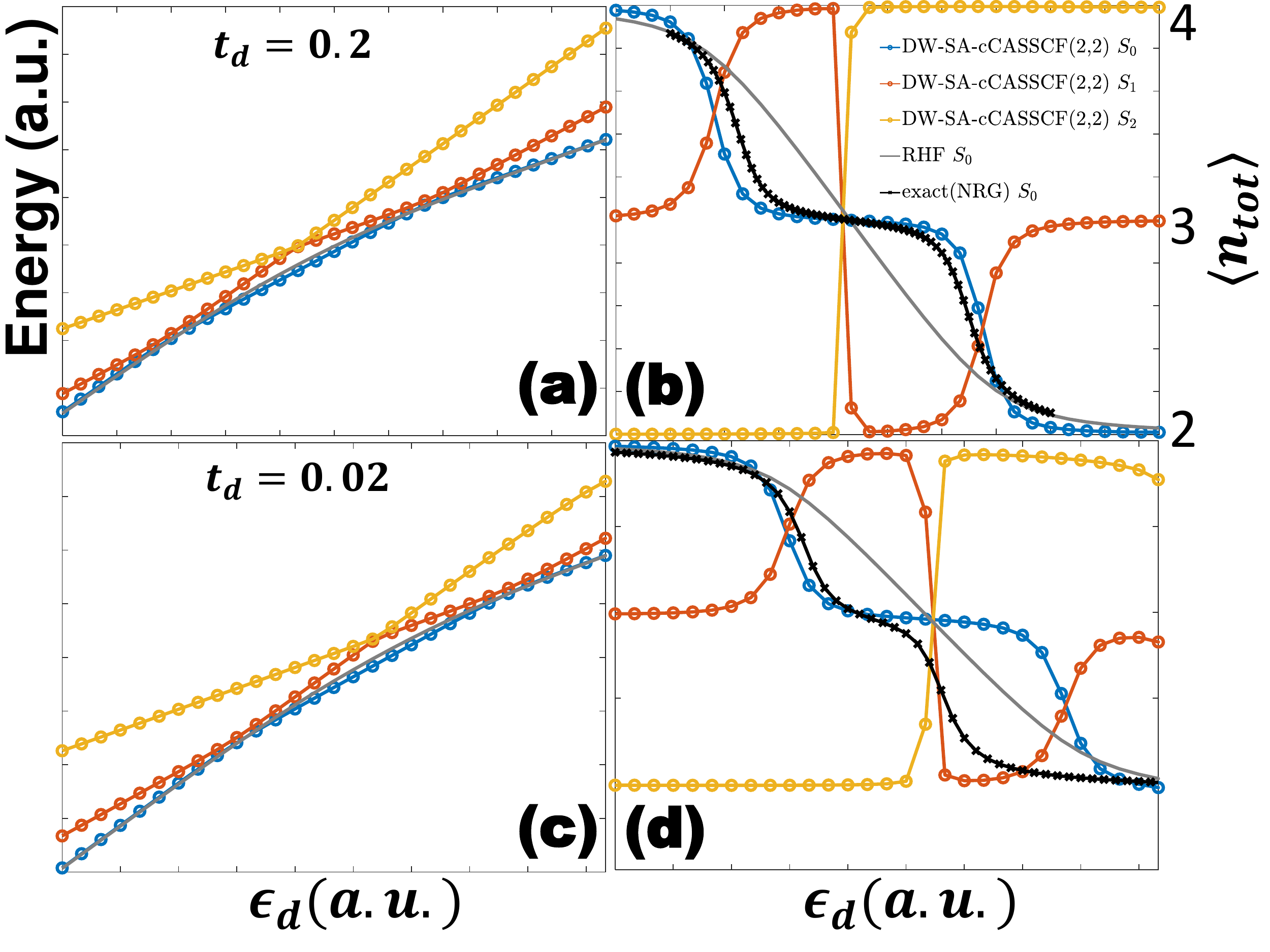}
\caption{The energies and the impurity electron population ($\expval{n_{tot}}$) of the 2-site Anderson impurity model {\em with} electron-electron repulsion for dynamical-weighted state-averaged constrained CASSCF(2,2) (DW-SA-cCASSCF(2,2)) with $\zeta=40$. (a) Strong-bonded case with $t_d=0.2$; (b) Weak-bonded case with $t_d=0.02$. Subfigures (a,c) don't include the harmonic nuclear potentials. On the right hand side, the black line is the exact adiabatic ground state electron population (calculated by numerical renormalization group, NRG). The blue line without the circle is the RHF energy and electron populations. In general, DW-SA-cCASSCF(2,2) performs well but the method fails to match  NRG in the weak-bonded $t_d=0.02$ limit for the ground state electron population (as a bigger active space is likely required).
The parameter set is $\epsilon_{d_1}=\epsilon_{d_2}, \Gamma=0.01, U=0.1$ with 101 metal states evenly distributed within the energy window $[-0.05, 0.05]$.}
\label{fig: 2site}
\end{figure}

Lastly, the results above might appear artificial in the sense that there is only one impurity orbital. After all, according to the constraint in Eqn. \ref{eqn:constraint1}, there is much less flexiblity that one will find in a model with more than one site (and multiple impurity orbitals). To that end, in Fig. \ref{fig: 2site}, we plot the energy and electron population results according DW-SA-cCASSCF(2,2) for a 2-site model {\em with}  electron-electron repulsion. We explore both the limit where the impurities are strongly-bound together (large $t_d$, Figs. \ref{fig: 2site} (a,b)) or  weakly-bound together (small $t_d$, Figs. \ref{fig: 2site} (c,d)). We explore a large range of possible impurity energies with two independent electron transfer events. 

For the strong-bound limit, in  Fig. \ref{fig: 2site}(a,b), we see that DW-SA-cCASSCF(2,2) again performs well, recovering reasonably accurate impurity populations over the regime where the impurity is occupied by two  or three or four electrons. However, in the weakly-bound limit ( Fig. \ref{fig: 2site}  (d)), we do find some qualitative errors.
While  DW-SA-cCASSCF(2,2) can capture an accurate ground state as the impurity population switches from four to three electrons, the electron population fails to match with NRG as the population switches from three to two electrons. In this regime, it would appear that there need to be several unpaired electrons, as we have both an electron transfer and a bond-breaking event simultaneously;  apparently, an active space of 2 electrons and 2 orbitals is not large enough  to fully account for such a process. In principle, one should be able to  improve these results either by increasing the active size or performing a configuration interaction on top of the CASSCF(2,2) reference states. These approaches will be taken up in a future paper.
%The goal of this letter is to report one possible simplest way to generate the ground and excited states for molecules near metal surfaces, so we will put aside the weak-bonded limit failure for now and apply our method to the non-bond-breaking processes in the future study. 

\section{Discussion: The Choice of Constraint}

% XXX Need 3 Paragraphs below:
% 1. Summary of what has been achieved, and what is still lacking.
% couplings
% choice of zeta
% choice of constraint
% equations 
% dicusssion of algorithm 
% benchmark of timings
% ab initio implementation
% We will address most of these will be in an upcoming paper.

Our results above have highlighted the power of a constrained dynamically weighted state-averaged CASSCF(2,2) calculation so as to model a molecule on a surface.  In order to run {\em ab initio} molecular dynamics on a metal surface, however, quite a few  more topics will need to be explored and made more efficient, as we will now discuss.

The first and most glaring omission in the present letter are the timings or computational cost of the present algorithm. For the most part, the present calculations were coded in a fairly inefficient manner, requiring on the order of minutes for completion. Future work will need to benchmark these timings and investigate the best optimization routine so as to minimize them.
Second, in order to run nonadiabatic dynamics at a metal surface (as opposed to the case of solution), one will require couplings to the metal surface (capturing electronic motion in the absence of nuclear motion). These couplings will be reported in a future publication.
Third, in this letter, we have not investigated how our results depend on the dynamical-weighting factor $\zeta$. This too will be reported in a future publication.
Fourth, in practice, it will be crucial to extract gradients, derivative couplings, and diabatic states from the present ansatz. All of these features will be essential if we are to eventually run {\em ab initio} simulations.

Fifth and finally, although we have shown in this letter why implementing a constraint can be crucial when applying CASSCF to the case of a molecule on a metals surface, we have not demonstrated that the constraint in Eq. \ref{eqn:constraint1} is actually optimal. In particular, one can easily imagine replacing Eq. \ref{eqn:constraint1} (where two active orbitals have some overlap with the impurity atomic orbitals ) with a fully localized constraint
( where we force each orbital to be either on the molecule or in the metal).
Mathematically, one can imagine replacing Eq. \ref{eqn:constraint1} (which we might call a partially localized constraint pcCASSCF(2,2) ) with the fully localized  constraint (which we might call a fully localized constraint fcCASSCF(2,2) ) in Eq. \ref{eqn:constraint2}: 
\begin{equation}
    \begin{aligned}
    \label{eqn:constraint2}
    \sum_{\mu\in\textbf{impurity}}\mel{t}{d_\mu^{\dagger}d_\mu}{t}=1 ;\\  \sum_{\mu\in\textbf{impurity}}\mel{u}{d_\mu^{\dagger}d_\mu}{u}=0
    \end{aligned}
\end{equation}
Differentiating between these two constraint frameworks is essential for understanding how best to understand and conceptualize how electron correlation functions a metal surface. After all, Eq. \ref{eqn:constraint2} is just about the simplest framework one can imagine implementing for modeling electron transfer--the equivalent of standard constrained DFT but for CASSCF. What are practical differences between these two constraints?

%At present, in this discussion section, we wish to discuss the role and the function of the constraint. Clearly, Eq. \ref{eqn:constraint1} is not unique and one can strict the constraint Eq. \ref{eqn:constraint1} by setting:
%\begin{eqnarray}
%    \sum_{\mu\in\textbf{impurity}}\mel{t}{d_\mu^{\dagger}d_\mu}{t}=\frac{a}{a+b} ;  \sum_{\mu\in\textbf{impurity}}\mel{u}{d_\mu^{\dagger}d_\mu}{u}=\frac{b}{a+b}
%\end{eqnarray}.
%This set of equations already satisfy Eq. \ref{eqn:constraint1} and if one sets $b=0$, it becomes Eq. \ref{eqn:constraint2}.

Now, it is important to emphasize that, in  the single-impurity case, these two constraints are equivalent. Suppose $\{t, u\}$ are two active orbitals solved by pcCASSCF(2,2). For any CASSCF solution, the wavefunction is invariant by the unitary rotation within the active space, which provides one degree of freedom (the rotation angle $\theta$) to formulate a set of new active orbitals $\{t', u'\}$:
\begin{equation}
    \begin{aligned}
        \label{eqn:rotate}
        \ket{t'}&=\cos(\theta)\ket{t}+\sin(\theta)\ket{u}\\
        \ket{u'}&=\sin(\theta)\ket{t}-\cos(\theta)\ket{u}
    \end{aligned}
\end{equation}
In single impurity case, the rotation angle $\theta$ can be chosen to satisfy\cite{onesite}: $|\braket{d}{t'}|^2=1$, where $d$ represents the impurity atomic orbital, and using the fact that $|\braket{d}{t'}|^2+|\braket{d}{u'}|^2=1$, it follows that $|\braket{d}{u'}|^2=0$. Thus, for a single state impurity, fcCASSCF(2,2) and pcCASSCF(2,2) are equivalent. However, this statement no longer holds for multi-impurity case, e.g. two-impurity case, because we are not guaranteed to find a $\theta$ to satisfy\cite{twosite}:
\begin{equation}
    |\braket{d_1}{t'}|^2+|\braket{d_2}{t'}|^2=1
\end{equation}

\begin{figure}[H]
\centering

\begin{subfigure}[t]{0.45\textwidth}
\centering
\hspace*{-0mm}\includegraphics[width=1.0\linewidth]{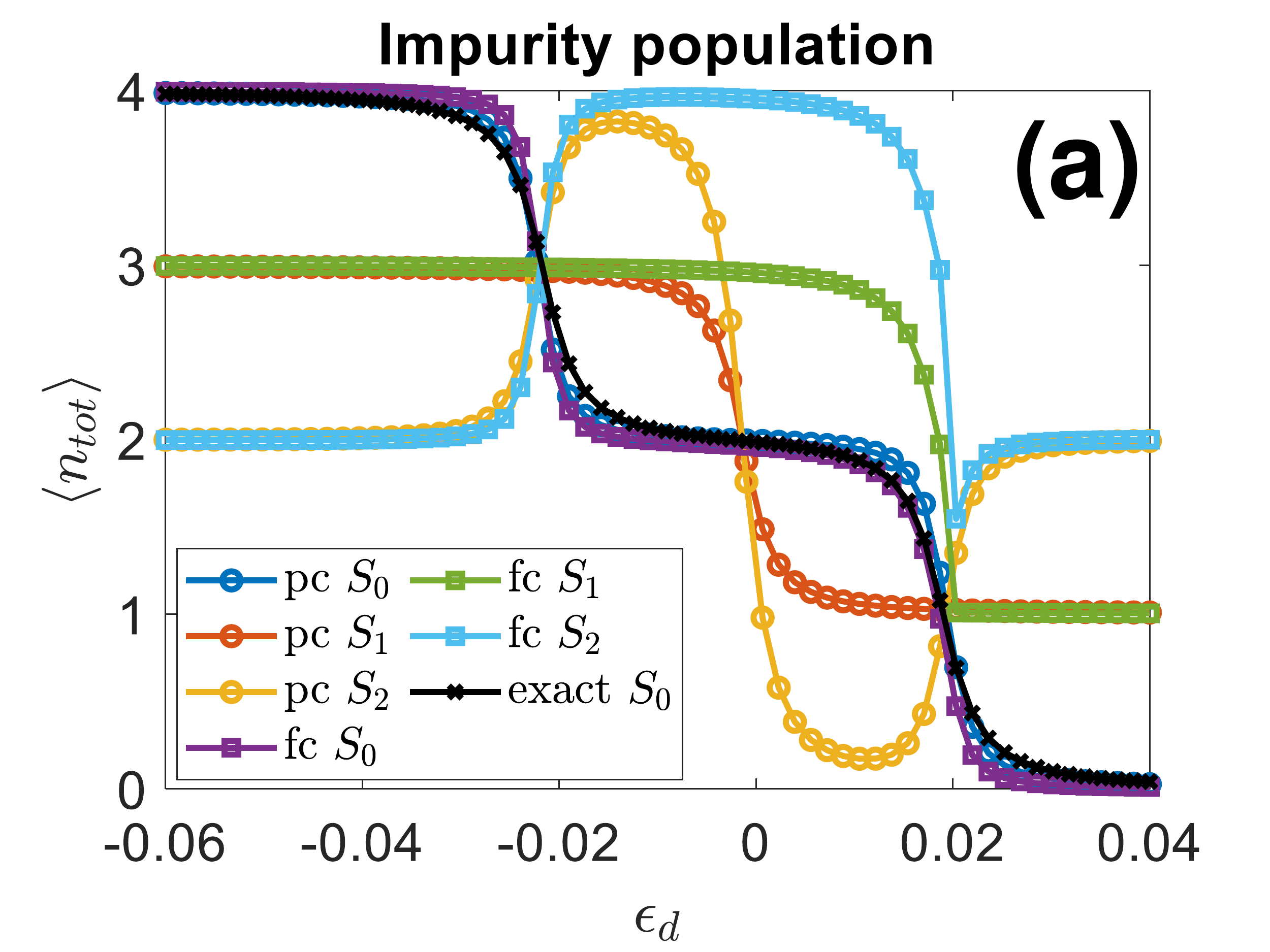}
\end{subfigure}
\begin{subfigure}[t]{0.45\textwidth}
\centering
\hspace*{-0mm}\includegraphics[width=1.0\linewidth]{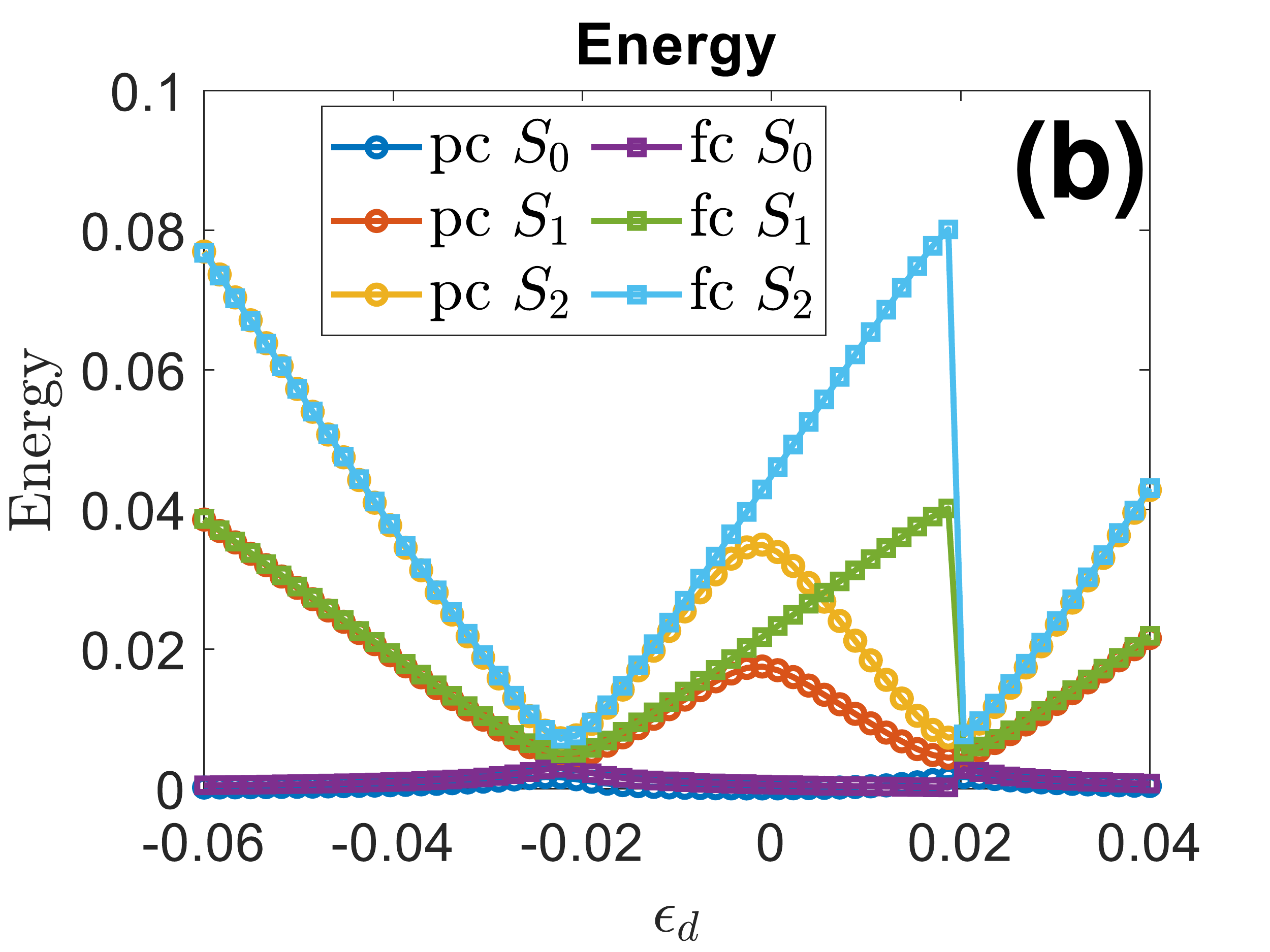}
\end{subfigure}

\caption{Comparison of partially localized constraint (pc for DW-SA-pcCASSCF(2,2) ) and fully localized constraint (fc for DW-SA-fcCASSCF(2,2) ). (a) Electron populations; (b) Energies (relative to exact). Note that DW-SA-fcCASSCF(2,2) suffers the multiple solution problem (the cause of the discontinuity at $\epsilon_d=0.02$); the partial constraint is smoother and more robust. The parameter set is:$\epsilon_d=\epsilon_{d_1}=\epsilon_{d_2}, \Gamma=0.01, U=0, t_d=0.02, \zeta=40$ with 31 metal states evenly distributed within the energy window $[-0.015, 0.015]$.} 
\label{fig:pcfc}

\end{figure}

To that end, we have coded up and compared 
Eq. \ref{eqn:constraint1} vs. Eq. \ref{eqn:constraint2}.
%We will continue this discussion and show this difference more rigorously according to 
%a single-particle perspective
% eigenvalues of active orbitals 1PDM
%in the next paper.
Our results for the two-site model are shown in 
 Fig. \ref{fig:pcfc}.
 In Fig. \ref{fig:pcfc}(a), we find that the pcCASSCF(2,2) results are inversion symmetric around the point $(\epsilon_d,\expval{n_{tot}})=(0,2)$. This inversion symmetry can be understood as a diabatic crossing between the diabat($\expval{n_{tot}}=3$) and the diabat($\expval{n_{tot}}=1$) for $S_1$ (or between the diabat($\expval{n_{tot}}=4$) and the diabat($\expval{n_{tot}}=0$) for $S_2$); pcCASSCF(2,2) captures  these transition smoothly. By contrast, we find that the fcCASSCF(2,2) results are not inversion symmetric and there is a discontinuity at $\epsilon_d=0.02$, indicating that there can be multiple solutions in the regime $\epsilon_d \in (-0.025,0.02)$. The energetic results shown Fig. \ref{fig:pcfc}(b) also support this observation.
In the end, the evidence would suggest that  fcCASSCF(2,2) will suffer a more serious multiple solution problem (and exhibit more discontinuities) than does pcCASSCF. Thus, we must conclude  that the latter is indeed a better starting point than the former as far as running dynamics.
 
% Clearly, one is better than other. In the end, it would appear, we have the right algorithm.

% In the discussion section, we will discuss the choice of two constraints on the two active orbitals. 
% \begin{itemize}
%     \item A looser and more flexible constraint (see the Eq. \ref{eqn:constraint1}), which requires two active orbitals have some overlap with the impurity atomic orbitals. We denote it as partially localized constraint ( pcCASSCF(2,2) );
%     % $\sum_{\mu\in\textbf{impurity}}\mel{t}{d_\mu^{\dagger}d_\mu}{t}+\mel{u}{d_\mu^{\dagger}d_\mu}{u}=1$;
%     \item A more strict constraint, which forces one active orbital to be fully localized on the molecule and the other active orbital to be a pure bath orbital. We denote it as a fully localized constraint ( fcCASSCF(2,2) ): 
%     \begin{eqnarray}
%     \sum_{\mu\in\textbf{impurity}}\mel{t}{d_\mu^{\dagger}d_\mu}{t}=1 ;  \sum_{\mu\in\textbf{impurity}}\mel{u}{d_\mu^{\dagger}d_\mu}{u}=0
%     \end{eqnarray}
% \end{itemize}

\section{Conclusion}
In conclusion, in this letter, we have reported a modified CASSCF(2,2) method, dynamically-weighted state-averaged constrained CASSCF(2,2) (DW-SA-cCASSCF(2,2)), which can generate smooth crossings and demonstrably accurate ground states   for simple molecules near metal surfaces; inclusion of a constraint is essential for picking out the relevant manifold of states and dynamical weighting is key for ensuring the smoothness. While the accuracy of this approach will no doubt be limited when larger molecules (with more electron correlation) are studied, one can hope that this approach will be able to offer some qualitative accuracy at the least to a very difficult problem where few nonadiabatic effects have been investigated in detail (though see recent work of Levine {\em et al} for an interesting CAS calculation of dangling bonds on a silicon cluster)\cite{peng2018dynamics}.
One can also imagine expanding the present method to much larger active spaces in the spirit of heat bath configuration interaction self-consistent field (HCISCF)\cite{smith2022near}, where nuclear gradients are now available. 

Looking forward, two potential directions would be very exciting to explore in the future. First, armed with a set of well-defined ground state and excited states, one could very much like to  study nonadiabatic dynamics directly and observe how both nuclear motion and bath-induced eletroncic relaxation in the metal contribute (and potentially compete) when molecules vibrate in the presence of a continuum of electronic states. To that end, we are in the process now of implementing the fewest-switches-surface-hopping with electron-relaxation (FSSH-ER) algorithm \cite{jin2021nonadiabatic} to study the Anderson model {\em with} the electron-electron correlation effect, which represents a more realistic model for molecules near metal surfaces.  Second, in the future, it will be very interesting to adapt the present DW-SA-cCASSCF(2,2) method %in a semi {\em ab initio} model where a real molecule talks to a tight-binding metal surface
to an {\em ab initio} electronic structure package, so that we can make concrete predictions about realistic molecules that are bound to a metallic surface and then oxidized or reduced. If successful, this approach should  make contact with a range of both inner and outer sphere  electrochemical experiments, allowing us to probe questions about the adiabaticity or nonadiabaticity of such processes ; answers to these questions as of yet are still largely unknown.

\begin{acknowledgement}

This work was supported by the U.S. Air Force Office of Scientific Research (USAFOSR) under Grant Nos. FA9550-18-1-0497 and FA9550-18-1-0420. We thank the DoD High Performance Computing Modernization Program for computer time.
We thank Xuezhi Bian for helpful discussions regarding  state-averaged CASSCF. We also thank Yanze Wu for helpful discussions regarding the nuclear degrees of freedom in the one-site Anderson Impurity model {\em with} electron-electron repulsion. 
\end{acknowledgement}

\begin{suppinfo}

This will usually read something like: ``Experimental procedures and
characterization data for all new compounds. The class will
automatically add a sentence pointing to the information on-line:

\end{suppinfo}

\bibliography{achemso-demo}

\begin{mcitethebibliography}{78}
\bibitem[one()]{onesite}
If we parametrize $x=\braket{d}{t}$, $y=\braket{d}{u}$, then we have three
  equations (1 coming from Eq. \ref{eqn:constraint1} and 2 coming from Eq.
  \ref{eqn:rotate}) to solve three variables $x,y$ and $\theta$.\relax
\mciteBstWouldAddEndPunctfalse
\mciteSetBstMidEndSepPunct{\mcitedefaultmidpunct}
{}{\mcitedefaultseppunct}\relax
\EndOfBibitem
\bibitem[two()]{twosite}
If we parametrize $x=\braket{d_1}{t}$, $y=\braket{d_1}{u}$,
  $z=\braket{d_2}{t}$, $w=\braket{d_2}{u}$ and it will be three equations
  solving 5 variables $x,y,z,w$ and $\theta$.\relax
\mciteBstWouldAddEndPunctfalse
\mciteSetBstMidEndSepPunct{\mcitedefaultmidpunct}
{}{\mcitedefaultseppunct}\relax
\EndOfBibitem
\end{mcitethebibliography}

\end{document}